\documentclass[preprint2,usenatbib]{aastex}

\newcommand{\amm}{NGC~6791}

\newcommand{\gsim}{\;\lower.6ex\hbox{$\sim$}\kern-7.75pt\raise.65ex\hbox{$>$}\;}
\newcommand{\lsim}{\;\lower.6ex\hbox{$\sim$}\kern-7.75pt\raise.65ex\hbox{$<$}\;}

\shorttitle{Metallicity of NGC 6791}
\shortauthors{R. Gratton et al.}

\begin{document}

\title{The metallicity of the old open cluster NGC 6791\altaffilmark{1}}
\altaffiltext{1}{ 
Based on observations made with the Italian Telescopio Nazionale Galileo (TNG)
operated on the island of La Palma by the Fundaci\' on Galileo Galilei of the
INAF (Istituto Nazionale di Astrofisica) at the Spanish Observatorio del Roque
de los Muchachos of the Instituto de Astrofisica de Canarias}

\author{Raffaele  Gratton\altaffilmark{2}, 
Angela Bragaglia\altaffilmark{3}, Eugenio Carretta\altaffilmark{3}  
and Monica Tosi\altaffilmark{3}}
\altaffiltext{2}{
INAF-Osservatorio Astronomico di Padova, vicolo Osservatorio 5, I-35122 Padova,
Italy}
\altaffiltext{3}{
INAF-Osservatorio Astronomico di Bologna, via Ranzani 1, I-40127 Bologna,
Italy}
\email{gratton@pd.astro.it, angela.bragaglia@bo.astro.it, 
eugenio.carretta@bo.astro.it, monica.tosi@bo.astro.it}
\date{}

\begin{abstract}
We have observed four red clump stars in the very old and metal-rich open
cluster  \amm \ to derive its metallicity, using the high resolution
spectrograph SARG mounted on the TNG. Using a spectrum synthesis technique we
obtain an average value of [Fe/H] = +0.47 ($\pm 0.04$, r.m.s. = 0.08) dex.
Our method was tested on $\mu$~Leo, a well studied metal-rich field giant. We
also derive average oxygen and carbon abundances for \amm \ from synthesis of
[O {\sc i}] at 6300~\AA \ and C$_2$ at 5086~\AA, finding [O/Fe] $\simeq -0.3$
and [C/Fe] $\simeq -0.2$.

\end{abstract}


\clearpage 

\section{Introduction}

The determination of the abundances in stars of different age and Galactic 
location is one of the basic tools to interpret the chemical evolution of the
Milky Way disk. Galactic open clusters are particularly well suited to this
purpose (e.g., \citealt{friel95}) since they reach ages as old as the disk
itself, cover a large range in metallicities and ages, are observed in  regions
of the Galactic disk likely characterized by different star formation
histories, and their distances and ages can be determined with a precision not
reachable for field stars, except for the nearest ones.

We are presently studying in an accurate and homogeneous way a significant
sample of open clusters (\citealt{bt06} and references therein): reliable
distances, reddenings and ages are derived from photometry with the synthetic
color-magnitude diagrams technique \citep{tosi91}, and will be combined with
metal abundances from high resolution spectroscopy to provide robust
constraints on the current and past disk properties.

We have already presented the detailed chemical abundances of a few old open
clusters  \citep{bragaglia01,c04,c05}. To our sample we now add \amm \ which,
with an age of about 9-10 Gyr,  is one of the oldest open clusters of our
Galaxy and has super solar metallicity (e.g., \citealt{pg98};
\citealt{chaboyer99}; \citealt{stetson03}; \citealt{carney05};
\citealt{king05}). This cluster, almost as old as the Galactic disk, is of
paramount importance to study the time evolution of the disk properties. Apart
from its age, NGC 6791 is interesting for his peculiar Horizontal Branch (HB),
mostly composed by red stars, but with a (unusual) blue tail
\citep{ku92,liebert94}. Its study could be relevant for a number of issues,
like e.g., the UV upturn in elliptical galaxies, the HB morphology and its
connection with mass loss in globular clusters.

Despite its peculiarities and its importance for Galactic formation studies, 
only one detailed work dealing with the chemical pattern in \amm, based on
modern fine abundance analysis and high resolution spectroscopy, has been
published so far: \cite{pg98} analyzed  the coolest and brightest (at $V =
15.0, ~B-V = 0.48$)  blue  HB star. Besides this case, really high S/N,  high
resolution spectra can be obtained in acceptable exposure times only for the
brightest, hence cooler giants in this cluster. At the very large metallicity
of \amm \ these pose a great challenge to the observers (see next Section) and
great care has to be taken to ensure the reliability of the analysis of their
extremely crowded spectra.

This paper is organized as follows: Sect. 2 presents our data; atmospheric
parameters and iron abundances from spectrum synthesis are described in Sect.
3; Sect. 4 provides the abundances of carbon and oxygen;  a discussion and a
summary are given in Sect. 5.

\section{Observational material}

\begin{figure*}
\includegraphics[bb=20 140 565 680, clip, angle=0, scale=.85]{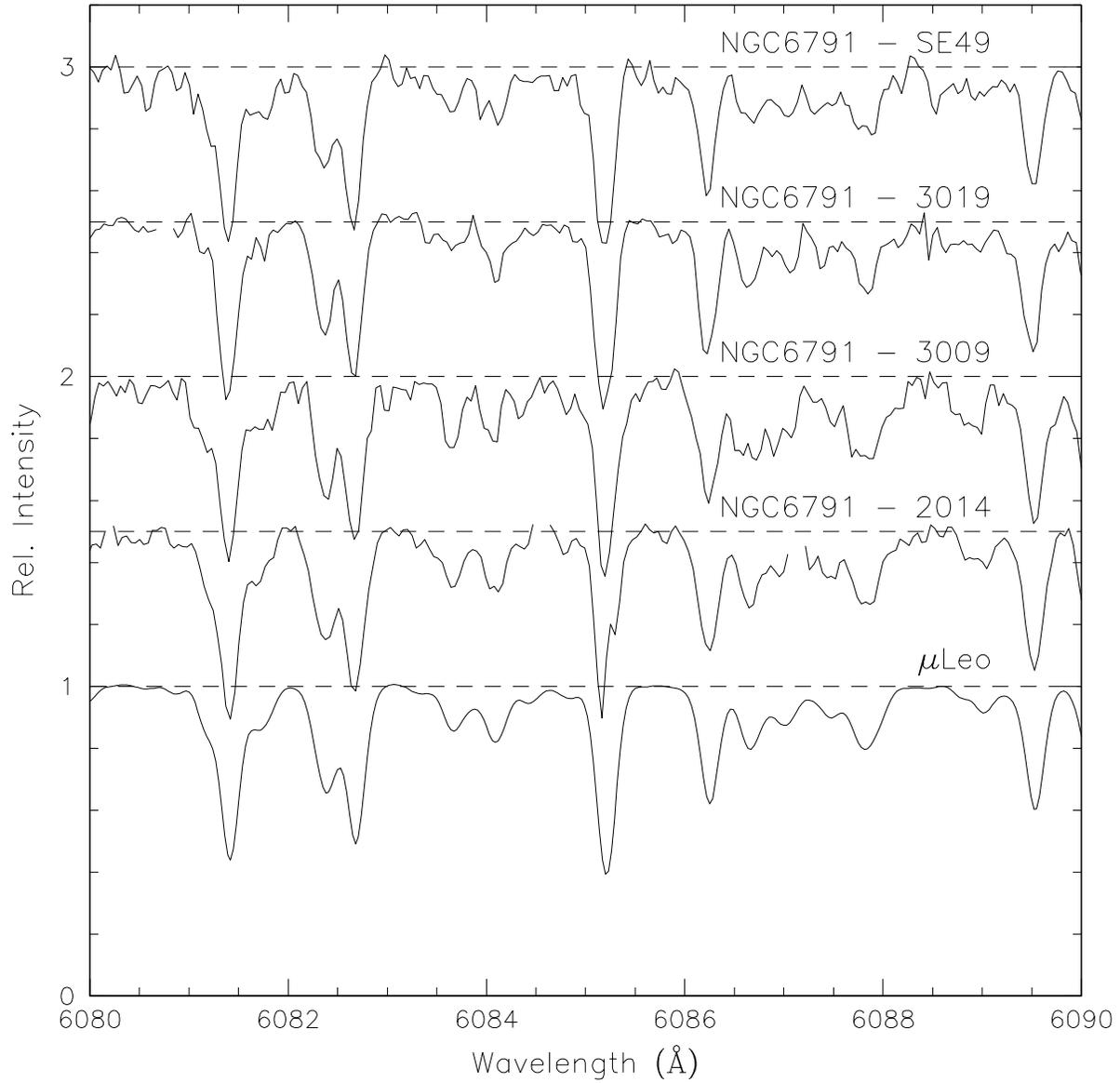}
\caption{Small portion of the spectra of our target stars compared with the
one of $\mu$ Leo, degraded to the same resolution.
}
\label{fig0}
\end{figure*}

The spectra of the giants of \amm \ are extremely rich of lines, due to the
rather cool temperature and the high metal content. To alleviate the analysis
problems, we focused our attention on (fainter) stars on the red clump, which
are warmer than the red giants. Even these spectra are actually at least as
rich in lines as that of the canonical very metal-rich giant $\mu$~Leo (see
Figure~\ref{fig0}). The long debate on the appropriate abundance to be
attributed to this star (see e.g. \citealt{gs90}) emphasizes the difficulties
of the derivation of correct abundances from the very line-rich spectra of cool
giants.

We chose our targets among the stars listed as cluster members by \cite{friel89}
on the basis of
their radial velocities (RVs). They are all confirmed members 
by our own measurements. All spectra were obtained with the high resolution
spectrograph SARG  mounted at the Italian National Telescope Galileo (TNG) on
Canary Islands. The resolution is $R = 29,000$ and the wavelength coverage is
4620 - 7920~\AA. Individual spectra have been obtained (mostly in service mode)
from November 2001 to September 2005, with exposures ranging from 1 to 1.5
hours each. For each  star, the total exposure time ranges from 4.5 to 8
hours. All spectra have been reduced using a standard IRAF\footnote{IRAF is
distributed by the National Optical Astronomical Observatory, which are
operated by the Association of Universities for Research in Astronomy, under
contract with the National Science Foundation} procedure for bias and flat
field correction, spectra extraction, and wavelength calibration; the
individual RVs were derived and the spectra were shifted to zero RV and
averaged. Table \ref{tab1} lists the identifications, coordinates and
magnitudes for the stars, together with the heliocentric RVs obtained averaging
all individual ones for each stars. Also shown is the S/N of the summed spectra
measured near 6000~\AA \ by comparing the spectrum of the stars with those of
$\mu$~Leo (see below), that can be considered virtually noiseless for the
purposes of this comparison.

To cross-check our method of abundance derivation, we also analyzed  a spectrum
of $\mu$~Leo with the same procedure. This star was selected because 
it has stellar parameters and iron abundances
similar to those expected in the NGC 6791 sample. 
The spectrum of $\mu$~Leo was acquired using
the FEROS spectrograph at the ESO 1.5~m telescope at La Silla. The original
spectrum has a resolution of $R\sim 48,000$; however, in order to make the
analysis as similar as possible to that of the stars of \amm, this spectrum was
degraded at the same resolution of our SARG spectra. As evident from
Figure~\ref{fig0}, at this resolution the spectrum of $\mu$~Leo is indeed very
similar to those of the stars of \amm, except for the higher S/N, and very
subtle differences in the line strengths. The fact that we are dealing with a
different, lower resolution spectrum than the one analyzed by \cite{gs90} and
adopt a different solar model explain the slight differences in the
results with that paper (see Sect. 3.4).

\section{Abundances in NGC6791 from synthesis of Fe lines}

In \amm \ even the spectra of red clump stars are so crowded that we deemed the
traditional analysis based on equivalent widths measure not entirely reliable,
due to the large inherent uncertainties in both the location of the continuum
level and the presence of blends (see \citealt{gs90} and  \citealt{sr00} for a
similar approach in the case of spectra of much higher resolution and S/N of
$\mu$~Leo). 

We then derived iron abundances for our program stars by comparing
the observed profiles for a number of Fe lines to syntheses of small spectral
regions (typical width $\sim 2.5$~\AA) around the chosen lines. This procedure
allows to take fully into account the presence of blends. Furthermore, the
correct positioning of the continuum level is much less a problem, since we
may compare directly the highest points of the spectrum in the observed
regions with those present in the synthetic spectra, insofar we trust the line
lists used in our analysis. These lists were built taking appropriately into
account both the inclusion of all relevant lines and the quality of the $gf$
values, carefully discussed in \cite{gratton03}.

These (extensive) line lists were obtained after
careful comparisons with both the spectrum of the Sun and of HR~3627. This
is a cool ($\sim 4200$~K), very metal-rich ([Fe/H]$\sim +0.3$) star. With
this combination of parameters the strength of lines in HR~3627 is typically
similar to (or even stronger than) that of the red clump stars of \amm. We are
fully confident that a line list well matching the spectrum of this star 
also provides sensible results for our program stars. This lengthy preparatory
work on HR~3627 was based on excellent observational material: its spectrum, 
acquired with SARG, has both very high S/N ($>400$) and resolution ($R\sim
150,000$), so that even extremely faint possible contaminants could be
detected and included in the line lists. 
We selected a number of Fe {\sc i} and Fe {\sc ii} lines that were free from
nearby strong features. We restricted to the spectral range from about 5500 to
7000~\AA,  to avoid the low response of the spectrograph in the blue region and
fringing or severe telluric contamination redward of the upper wavelength
limit. Around each lines (within $\pm $2~\AA) we extracted   lines of neutral
and singly ionized atomic species from the Kurucz database (\citealt{kur95b}).
We also included molecular lines,  in particular of contaminant CN and
hydrides. Lines unaccounted for in the Kurucz database and/or in the solar
tables (\citealt{moore}) were  assumed to be Fe {\sc i} lines with an
excitation potential of 3.5~eV.  The transition probabilities of the Fe lines
we wanted to synthetize were left untouched with respect to the line list we
used in the $EW$ analysis; the $gf$ values of the nearby lines were adjusted
one by one by matching the high resolution spectrum of HR 3627. 
For more
details about HR 3627 and these line lists, see \cite{c04}.

The synthetic spectra were obtained using model atmospheres extracted from the
grid of \cite{kur95a}. For consistency with other papers analyzing stars in open
clusters (see e.g. \citealt{c04}), the models considered in this paper are
those with the overshooting option included.

The fitting of the synthetic spectra to the observed ones was done by
eye. There is of course some arbitrariness in the eye fitting, since different
weights can be attributed to the line cores or wings. However, we found that
our eye fitting gives the same average abundances, but much less line-to-line
scatter, than a fitting based on more objective criterions, such as a least
square fitting to the data. The reason for the smaller scatter 
likely reflects a better estimate of the appropriate level
of the local continuum, which is a free parameter in the fitting, and it is a
critical issue when determining abundances.

The line parameters (oscillator strengths, damping broadening) were obtained
using the same precepts adopted in \cite{gratton03}. The same line
parameters and  microturbulent velocity were adopted for all stars.

\subsection{Atmospheric Parameters}

Effective temperatures ($T_{\rm eff}$) and surface gravities ($\log g$) were 
obtained from the photometry, using $B, V$ values from
\cite{montgomery94}\footnote{
but other photometries are almost identical, see e.g. \cite{stetson03}}
 and $K$ magnitudes from 2MASS \citep{cutri03}. We derived the $T_{\rm eff}$'s
from the $V-K$\ colors assuming the calibration by \cite{alonso99} and a
reddening of $E(B-V) = 0.15$, in the middle of literature determinations. We
prefer not to use temperatures from the $B-V$\ colors because of their strong
dependence on metal abundance. We have found a posteriori that the values of
the effective temperatures corresponding to the $B-V$\ colors agree well with
those  from the $V-K$\ colors for the metal abundance of \amm \ determined in
the present analysis. Also, we prefer not to use temperatures derived from line
excitation because for the lines considered here there is a quite strong
degeneracy between effective temperatures and microturbulent velocities (see
below). 

The calibration of the colour-temperature relations of \cite{alonso99} was used
for consistency with the analyses we are performing of several other open
clusters. It should be noticed that their calibration only extends up to
[Fe/H]=+0.2, and application to the stars of \amm \ requires an extrapolation.
However, the $V-K$ color index used in this paper is only very marginally
sensitive to metal abundance, so that errors in our temperatures cannot be
large.
Errors in our effective temperatures are mainly due to uncertainties in
the assumed reddening: an error of $\Delta E(B-V)=0.04$\ magnitudes, a
reasonable upper limit in the case of \amm, implies an error of $\sim 100$~K in
the assumed temperatures. For consistency, a similar approach was adopted for
$\mu$~Leo, with the $V-K$\ color from \cite{johnson66} photometry. Notice that
this $T_{\rm eff}$\ for $\mu$~Leo is slightly cooler (by 50~K) than that
derived by \cite{gs90} using the Infrared Flux Method.

Surface gravities were obtained from the location of the stars in the
color-magnitude diagram, assuming a distance modulus of $(m-M)_V=13.45$ - in
the middle of literature determinations - bolometric corrections from
\cite{alonso99}, and a mass of 0.9~$M_\odot$.  Most of the errors in the
surface gravities stem from uncertainties in the distance modulus: an error of
$(m-M)_V = 0.5$ mag, a reasonable upper limit, implies an error of
0.2~dex in the surface gravities. For $\mu$~Leo, we adopted the value of the
gravity given by \cite{gs90} using a variety of methods (equilibrium of
ionization for Fe, dissociation equilibrium for MgH, and pressure broadened
lines): this value is in fact almost identical to that of the NGC~6791 stars.

Microturbulent velocities ($v_t$) were obtained by eliminating trends of
abundances with respect to the expected line strength \citep{magain84}
$X=\log{gf}-EP \times \Theta_{exc}$, where $\log{gf}$ is the oscillator
strength of the lines, $EP$\ the excitation potential (in eV), and
$\Theta_{exc}=5040/(0.86 \times T_{\rm eff})$\ represents the approximate
temperature of the layers where most of the lines form. This implies that 
abundances are roughly independent of line strength. We considered here the
average values of the abundances derived for the individual lines of all the
stars of NGC~6791, in order to reduce the scatter and better evidentiate
possible trends. This means that the same value of the microturbulent velocity
was adopted for all stars. The same value was also adopted in the analysis of
$\mu$~Leo, and it is slightly smaller (by 0.15~km s$^{-1}$) than what 
\cite{gs90} and \cite{sr00} adopted. Uncertainties in these microturbulent
velocities are approximately of 0.08~km~s$^{-1}$ for a given effective
temperature; this is obtained by modifying $v_t$\ from its best value until the
slope becomes equal to its statistical error. Note however that there is a
strong correlation between microturbulent velocities and effective
temperatures: adopting temperatures 100~K higher, we would have derived
microturbulent velocities $\sim 0.13$~km s$^{-1}$ larger.

Final metallicities are obtained by interpolating in the \cite{kur95a} grid of
model atmospheres (with the overshooting option on) the model with the
proper atmospheric parameters whose abundance matches that derived from Fe
{\sc i} lines.

\subsection{Abundances from individual lines}

\begin{figure*}
\includegraphics[bb=20 140 565 680, clip, angle=0, scale=.85]{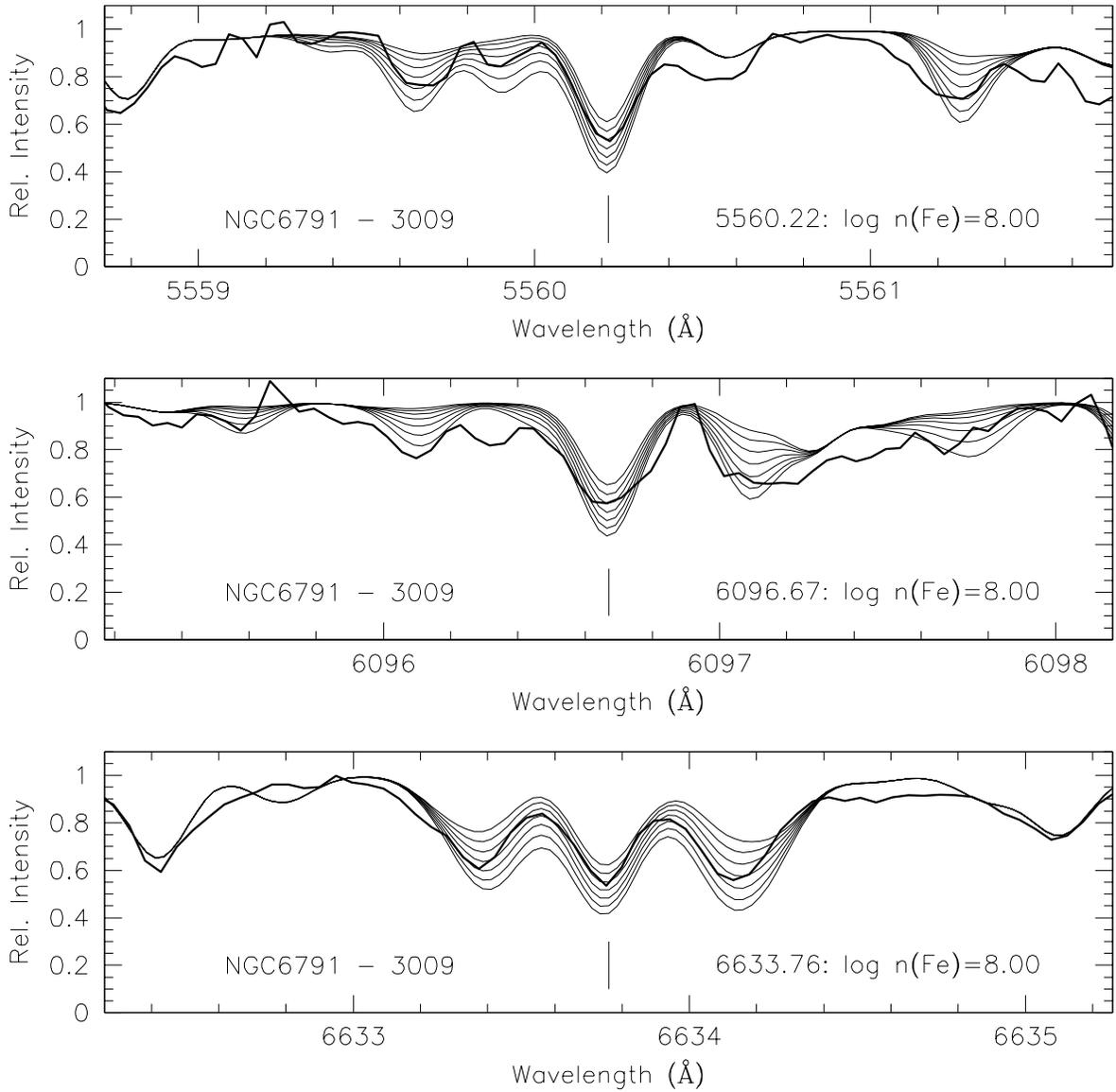}
\caption{Comparisons between the observed spectrum for a star in NGC 6791
(star 3009: thick line) and synthetic spectra (thin lines), for a few of
the spectral regions containing Fe lines considered in the present analysis.
}
\label{fig1}
\end{figure*}

Fig. \ref{fig1} shows an example of the quality of the fit of three of the
observed iron lines with synthetic spectra. The synthetic spectra have been
computed with atmospheric parameters appropriate for the star, and Fe
abundances of $\log n({\rm Fe})$=7.6, 7.8, 8.0, 8.2, 8.4, 8.6, and 8.8. The
lines considered in the analysis are marked with a dash in Fig. \ref{fig1}.
From these three comparisons we concluded for a best value of the Fe abundance
of $\log n({\rm Fe}) $=8.00. Note that other Fe {\sc i} lines falling in the
same spectral ranges of the program lines generally give Fe abundances in good
agreement with those indicated by the lines selected in our analysis, although
they were not actually used in the estimate of the best value for each star.

Table~\ref{t:abundances} gives abundances for Fe lines\footnote{These
abundances are by number; we use the usual spectroscopic notations: $\log
n({\rm A})$ is the abundance (by  number) of the element A in the scale where
$\log{n({\rm H})}=12$; [A/H] is the logarithmic ratio of the abundances of
elements A and H in the star, minus  the same quantity in the Sun.} 
as obtained from the comparison of observations with spectral synthesis. Lines
with expected line strength $X>-5.6$\ were not used in the analysis because
there is a trend for these lines to give too low abundances. This is likely a
reflection of the weight we gave in our abundance estimates to the cores of
these lines. These form at very tiny optical depths, where the adopted model
atmospheres are probably not adequate and deviations from LTE become important.
Anyhow, such lines are strongly saturated, so that they would not be good
abundance indicators. Since the four target stars are very similar to each
other, it is meaningful to average the results for the individual stars, and to
derive the average abundance provided by each line. This is given in the last
column of the table; the associated error bar is simply the dispersion of the
mean.

\begin{figure*}
\includegraphics[bb=20 140 565 680, clip, angle=0, scale=.85]{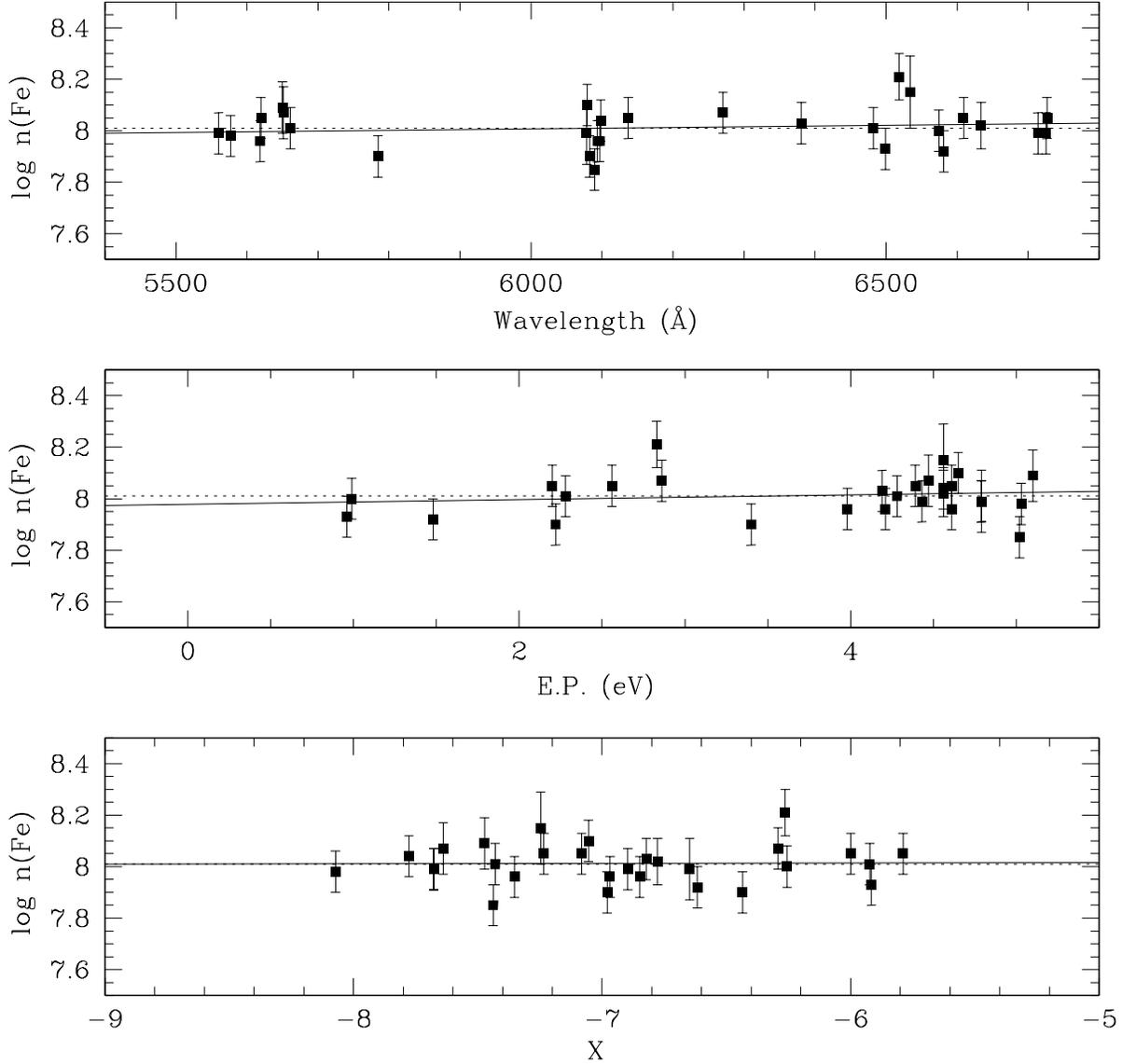}
\caption{Runs of the abundances obtained by averaging the results for
each individual Fe I line over the four program stars, with wavelength,
excitation potential $EP$, and the expected line intensity parameter
$X=\log{gf}-EP*5040/(0.86*T_{\rm eff})$. The error bars for the individual
lines are the maxima between the spread of the mean over the four stars,
and the r.m.s. of this quantity (0.08 dex). The average value of the
abundance for NGC 6791 is indicated by the dashed horizontal line, while the
formal best fit line based on Fe {\sc i} is indicated by the solid line. 
There is no significant trend in any of the panels of this figure.
}
\label{fig2}
\end{figure*}

Fig. \ref{fig2} shows graphically the lack of trends in this average Fe
abundances with wavelength, EP, and line intensity parameter X, thus
reinforcing the reliability of the derived abundances.

Table~\ref{t:summary} summarizes the abundances obtained for each star. The
second, third and fourth columns give the values adopted for the atmospheric
parameters ($T_{\rm eff}$, $\log{g}$, [A/H]), while the microturbulent velocity
is the same for all stars ($v_t$ =1.05 km s$^{-1}$). In Columns 5 to 10 we give
the average values of the abundances of Fe from neutral and singly ionized
lines, along with the number of lines used in the analysis and the r.m.s.
scatter of individual abundances. Finally, in the last two columns we give the
[Fe/H] values. The reference solar abundances adopted here are $\log n({\rm
Fe}) = 7.54$\ and 7.49 from neutral and singly ionized lines, respectively;
those are the values we obtained from a solar analysis compatible with the
present one (see \citealt{gratton03}).

The last line of Table~\ref{t:summary} gives the average abundance for the
cluster. The average Fe abundance from neutral lines is [Fe/H] = $+0.47\pm
0.04$\ (r.m.s. of individual stars equal to 0.07 dex). The line-to-line scatter
for each individual star is in the range 0.11-0.18 dex. This is dominated by
uncertainties still present in the location of the continuum level within the
small spectral windows considered in this analysis.

If we consider the average abundances for each line derived from the four
spectra, the line-to-line scatter is 0.08 dex, as expected by assuming that the
four results are independent of each other. For these average abundances,
there is a small, not significant, trend with line excitation:
$\Delta$[Fe/H]/$\Delta$EP = $0.009\pm 0.012$~dex/eV. This implies that
temperatures from line excitation are $38\pm 50$~K higher than those adopted
here. We deem this agreement as fully satisfying.

There is also a small - but again not significant - difference between the
abundances of Fe {\sc i} and Fe {\sc ii}: [Fe/H]{\sc i} $-$ [Fe/H]{\sc ii} =
$0.04\pm 0.07$~dex. This small difference would have canceled, had we chosen
temperatures about 50~K lower than adopted, or gravities about 0.15 dex
larger.

\subsection{Errors on the derived abundances}

Table~\ref{t:errors} gives the sensitivity of the derived abundances on
the assumptions on the atmospheric parameters of the program stars (listed in
Column 1). Column 2 gives the 
considered parameter variation, and Columns 3 and 4 the resulting variation in the abundances derived
from Fe {\sc i} and Fe {\sc ii} lines respectively. Column 5 lists the
estimated value for the systematic error in our analysis for each of the
parameters, and Columns 6 and 7 the corresponding uncertainties in the
abundances from Fe {\sc i} and {\sc ii} lines, respectively. 

The total errors listed on the bottom line of Table~\ref{t:errors} are obtained
summing quadratically the contribution of each source of error, including the
fitting error. This estimate of the total error is computed assuming a zero
covariance between the effects of errors in the atmospheric parameters. In
principle, this assumption is not strictly valid, since there are correlations
between different parameters.  In practice, the Fe abundance is mainly a
function of the adopted effective temperatures. In fact, adopting $T_{\rm
eff}$\ values  e.g. 200~K higher, we  would have obtained higher microturbulent
velocities (by  0.26~km~s$^{-1}$), and Fe abundances smaller by about 0.15~dex.
Notice that since the effect on microturbulent velocity is larger than the
direct effect of temperature variations, the abundances get smaller with
increasing model temperatures, at variance with the usual dependence for late
type stars. Unluckily, given the correlation between effective temperature and
microturbulent velocity, we could not derive reliable effective temperatures
from the excitation equilibrium.

There are quite strong arguments favoring the solution adopted throughout this
paper. In fact, had we adopted the hypothetical temperatures warmer by 200~K
mentioned above (that would be obtained assuming a reddening of $E(B-V) =
0.23$, a value near the upper limit in the  analyses of \amm), abundances from
Fe {\sc ii} lines would have resulted  much lower than those derived from Fe
{\sc i} lines, with an offset of 0.24 dex, rather than 0.04 dex. In principle,
it should be possible to compensate for such an offset by increasing the
surface gravities by 0.5 dex (up to values of $\log g \sim 2.8$).  However,
this would require a distance modulus of $(m-M)_V\sim $12.15,  incompatible
with the distances derived from the literature color-magnitude diagrams. The
absolute magnitude of core He-burning stars (i.e., the clump stars we are
observing) is expected to be $M_V\sim 1.2$\ according to the models by
\cite{girsal01}. This corresponds to a distance modulus of $(m-M)_V\sim
13.4-13.5$, in agreement with the value used in our analysis. Hence, an upper
limit for the effective temperatures of the program stars in \amm \ is about
100 K higher than the adopted values (corresponding to a reddening of $E(B-V) <
0.18$, and a microturbulent velocity of $v_t < 1.18$ km~s$^{-1}$). This implies
that the lower limit of the Fe abundance is [Fe/H]$>+0.39$ dex. Similar
arguments can be used to define a robust upper limit of [Fe/H]$ < +0.55$. 

Summarizing, the Fe abundance we derive for \amm \ is [Fe/H] = $+0.47\pm
0.04\pm 0.08$, where the first error value represents the random term, as
derived from the star-to-star scatter, and the second error value the
systematic errors, due to the assumptions on reddening, distance modulus, mass,
temperature scales, etc.

\subsection{A comparison with $\mu$ Leo}

To further assess the soundness of  the results of our spectrum synthesis
analysis, we also analyzed the well known, metal-rich field star $\mu$~Leo,
strictly using the same criteria.  The average Fe abundance we derived is
$\log{n({\rm Fe})}=7.92\pm 0.03$\ (29 lines, r.m.s.=0.15~dex) from neutral
lines, and $\log{n({\rm Fe})}=7.79\pm 0.04$\ (6 lines, r.m.s.=0.10~dex) from
singly ionized lines. These values correspond to [Fe/H]=$+0.38\pm 0.03$\ and
[Fe/H]=$+0.30\pm 0.04$\ respectively. The difference in the abundances provided
by neutral and singly ionized lines is only marginally larger than the internal
error, and is similar to the corresponding difference found for the stars in
NGC~6791.

This value of the Fe abundance of $\mu$~Leo can be compared with literature
estimates. We will consider only abundances based on very high quality spectra
($R\sim 100,000$, $S/N>200$) and comparisons with synthetic spectra, since use
of equivalent widths is not reliable for such line-rich spectra. \cite{gs90}
derived $\log{n({\rm Fe})}=7.97\pm 0.02\pm 0.15$\ (36 lines, r.m.s.=0.12~dex).
Within the internal errors, the present estimate of the abundance agrees with
that of \cite{gs90}. The [Fe/H] value given by \cite{gs90} is slightly lower
than that estimated in this paper ([Fe/H]=$+0.34 \pm 0.03$), due to the use of
a different solar abundance, derived using the \cite{hm74} model atmosphere,
rather than that obtained from the same grid considered for the red giants.
More recently, \cite{sr00} used a very similar technique and set of atmospheric
parameters, and obtained a lower value of $\log{n({\rm Fe})}=7.79\pm 0.03$\
(internal error). This value is about 0.13 dex lower than the present one and
the difference can be ascribed to the larger value of the microturbulent
velocity adopted by \cite{sr00} (1.22~km~s$^{-1}$\ rather than
1.05~km~s$^{-1}$: see Table~\ref{t:errors}).  These comparisons show that
systematic errors in our analysis are within the adopted error bar.

By comparing the abundances in NGC~6791 with those of $\mu$~Leo, we conclude
that the former has an Fe abundance larger than the latter by $0.09\pm 0.05$,
where the error here is simply the quadratic sum of the internal errors, since
the other uncertainties should cancel out in the differential abundance
analysis. This small positive difference well agrees with the eye impression
from Figure~\ref{fig0}.

\section{Abundances of oxygen and carbon}

We estimated the abundance of oxygen from the [O {\sc i}] line at 6300.31~\AA.
The second, weaker, [O {\sc i}] line at 6363.79~\AA \ is barely measurable in
our spectra, and the three O lines near 7775~\AA \ (that are not the best
suited for this kind of stars) fall in a region where strong interference
fringes make it difficult to extract accurate information from spectra.   

The 6300.31~\AA\ line often happens to be strongly affected by telluric
contamination. However, in the case of \amm, the cluster RV relative to Earth
motion makes  the  [O {\sc i}] line  non contaminated by telluric absorptions
in the majority of our spectra. Compromising between a slightly lower S/N and a
lower data  manipulation, we decided to average only the undisturbed spectra
for each star, and measure the O abundance from them.

\begin{figure*}
\includegraphics[bb=20 140 565 680, clip, angle=0, scale=.85]{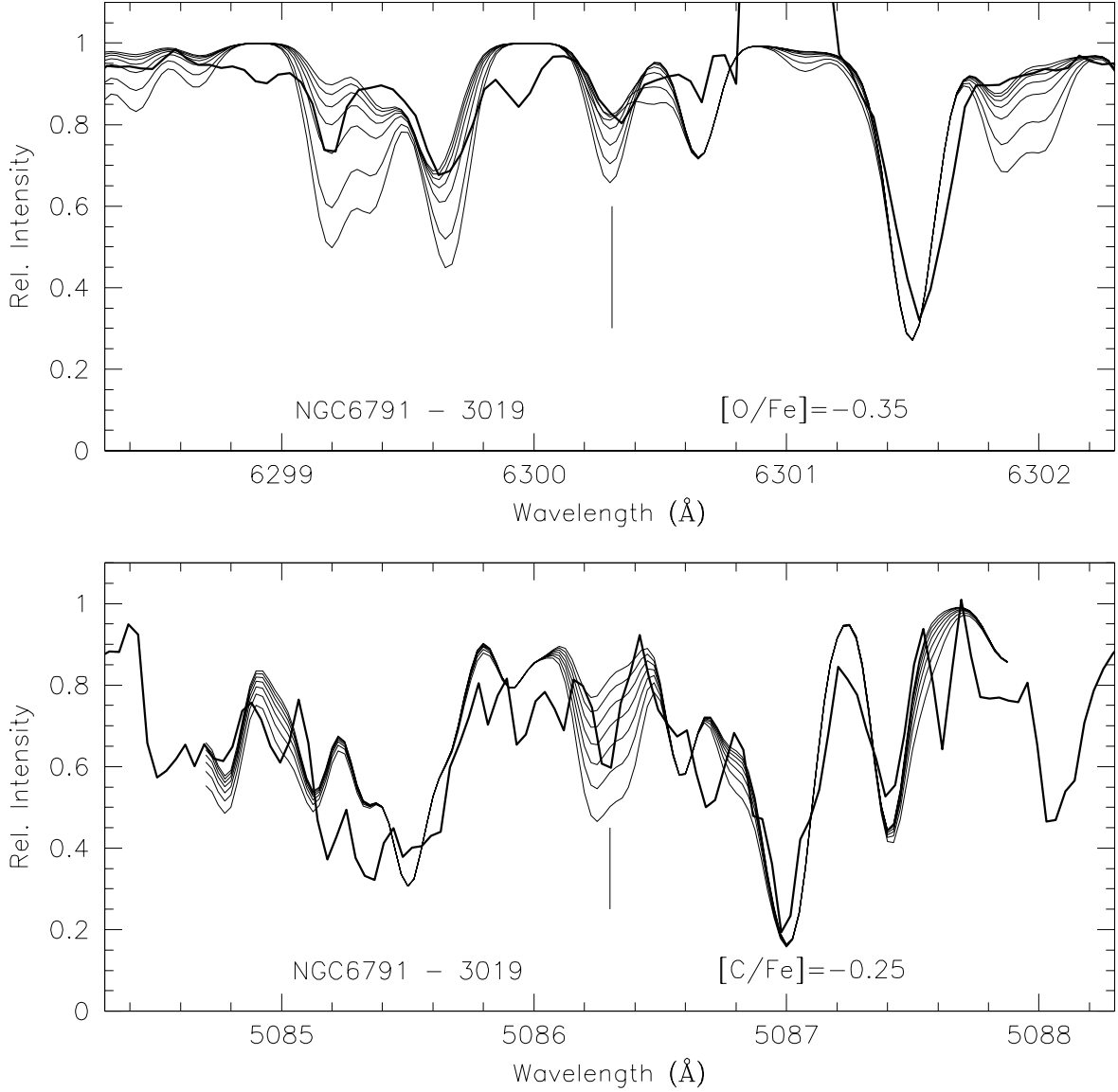}
\caption{Comparison between the observed spectrum for star 3019 in NGC 6791
(thick line) and synthetic spectra (thin lines) for the [O {\sc i}] line at
6300 \AA \ (upper panel) and the C$_2$ molecular feature at 5086 \AA (lower
panel). The synthetic spectra were computed assuming [N/Fe]=$-0.2$, and
$^{12}$C/$^{13}$C=8. In the upper panel we adopted [C/Fe]=$-0.25$, and
[O/Fe]=$-0.65, -0.55, -0.45, -0.35, -0.25, -0.15, -0.05$.  In the lower panel
we adopted [O/Fe]=$-0.35$, and [C/Fe]=$-0.40, -0.35, -0.30, -0.25, -0.20,
-0.15, -0.10$.}
\label{fig4}
\end{figure*}

In order to derive reliable O abundances, a careful synthesis of the  [O {\sc
i}] lines is necessary, including not only the contribution of the nearby Ni
line but also the relevant coupling with C abundances, and the contamination by
CN lines. While a full discussion of CNO abundances is deferred to a
forthcoming paper, here we preliminarily estimate the C abundances from the
spectral synthesis of the C$_2$\ molecular features at 5086~\AA. Fig.
\ref{fig4} shows an example of  matching of synthetic spectra for star 3019
in the forbidden [O {\sc i}] line and in the C$_2$ regions. The synthetic 
spectra were computed assuming [N/Fe]=$-0.2$, and $^{12}$C/$^{13}$C=8. The
latter value is adequate for low mass evolved giants, while the former one
provides  CN features in reasonable agreement with observations for the best C
abundance.

Combining the C and O abundances provided by these two abundance indicators, we
find average abundance ratios of [C/Fe] = $-0.23$ dex (a rather normal value
for clump stars), and [O/Fe] = $-0.32$ dex. The comparisons with synthetic
spectra clearly show that the O abundance cannot be much larger than this
value. It cannot be much lower either, otherwise the features due to C-bearing
molecules would be much stronger (the spectra of the stars clearly shows that
O$\gg$C).

As a comparison, \cite{gs90} derived for $\mu$~Leo [C/Fe] = [O/Fe] = $-0.15$
dex. \amm \ is then slightly more deficient in both O and C with respect to
$\mu$~Leo, along the standard trend of more pronounced deficiency of these
elements with increasing metal abundance (see \citealt{bensby05};
\citealt{andedv})

\section{Discussion and summary}

The first detailed study of \amm \ of which we are aware of is by
\cite{kinman65}, who published a photographic color-magnitude diagram and
derived information on membership and intrinsic colors (i.e., reddening) from
low resolution spectra.

There is a general agreement that this cluster is at about the same
Galactocentric distance as the Sun, is about twice as old, and about twice as
metal-rich. Nevertheless, there are significant differences between properties
derived by different authors: $(m-M)_0 \sim 12.6 - 13.6$, $E(B-V) \sim 0.09 -
0.23$, age $\sim 7 - 12$ Gyr, [Fe/H] $\sim 0.1 - 0.4$ (for a recent review see
\citealt{stetson03}). \amm \ has also been observed with the ACS on board the
Hubble Space Telescope (\citealt{king05}; \citealt{bedin05}), reaching almost
the hydrogen-burning limit on the main sequence and defining the White Dwarfs
cooling sequence.

Spectroscopic studies of \amm \ are fewer, but present interesting results.
Since the pioneering work by \cite{st71}  the cluster has been recognized to
be more metal-rich than the Sun; they found  [M/H] = +0.75, but with a method
that attributes [M/H] $\simeq$ +0.6 to NGC 188 and M 67, both presently
known to have nearly solar metallicity.

More recently, the red HB (that we call red clump) has been studied by low
resolution spectroscopy by  \cite{hufnagel95}; they wanted to investigate
possible correlations between CH and CN band strengths and found none, at
variance with similar studies on globular clusters.

\cite{wj03} took low resolution spectra of 23 K giants in \amm \ and derived,
using Lick/IDS indices, [Fe/H] = $+0.320 \pm 0.023 \pm \sigma_{sys}$, where
$\sigma_{sys}$, the systematic error attached to their metallicity scale, is
unknown. We have only one star in common (3019, their R25) for which the
quoted abundance is [Fe/H] = +0.341.

All the four stars analyzed here were observed at low resolution by
\cite{friel89}, where we adopted identification and membership status from.
The same group \citep{friel93} later published their metallicities: [Fe/H]
values are $+0.32 \pm 0.28$, $+0.23 \pm 0.33$, $+0.39 \pm 0.15$, $+0.41 \pm
0.21$ for stars 2014, 3009, 3019, and SE49, respectively. These values (that
would give an average metallicity of about +0.34) do not differ much from our
new ones, considering that the scale used by \cite{friel93} gives on average
metal-poorer values than others. The final result of their analysis depends  on
the calibration of indices using standard stars with metallicity derived  by
several different sources. Due to a recent revision of their abundance scale 
\citep{friel02}, these abundances were lowered to  $+0.12 \pm 0.13$, $+0.11 \pm
0.13$, $+0.15 \pm 0.11$ and $+0.16 \pm 0.11$ for the same stars;  for their
total sample of 39 stars they derived an average metallicity of +0.11,  $\sigma
= 0.10$  dex.  

To date, the only published analysis based on high resolution spectroscopy
remains that by \cite{pg98}. They obtained 4 hours of integration at the 4~m
Mayall telescope at Kitt Peak on a blue HB star that is a cluster member both
by proper motion and RV. The summed spectrum has S/N and resolution lower than
ours (S/N $\sim$ 30 per pixel, R = 20000), but is less affected by blends
because of the much higher temperature (about 7300 K). They employed both $EW$s
(for Fe and atmospheric parameters determination) and spectrum synthesis (for
all the other measured elements). They discussed the use of different scales,
e.g. for $\log gf$'s and solar reference abundances. On the scale they chose to
adopt, their star has [Fe/H] = +0.37 ($\pm 0.10$) dex; since their solar Fe is
$\log n$ = 7.67 and ours is 7.54, their value corresponds to [Fe/H] = +0.50 on
our scale.

Taken at face value, the agreement with our results appears excellent, but it
doesn't take into account many differences like e.g., $\log gf$'s or model
atmospheres and the like.  For instance, we have 4 Fe lines in common  with
\cite{pg98}, 1 of Fe {\sc i} and 3 of Fe {\sc ii} (see Table \ref{t:abundances}
and their table 1), and the average difference in $\log gf$'s is about 0.29
(our values minus theirs). They also measured the abundances of several
elements by means of extensive spectrum synthesis, finding strong
overabundances with respect to the  solar ratio for N and Na (+0.5 or +0.4) and
slightly lower ones for Mg and Si (+0.2) or Eu (+0.1). All the other elements,
including C and O, display solar-scaled abundances. In our analysis we have
measured so far only C and O abundances, finding for them sub-solar ratios. A
complete comparison between the two analyses is beyond the purpose of the
present paper, also because the chosen targets are quite different in
properties, but it will be done once we have measured all other elements.

Summarizing, we have derived the metallicity of \amm \ analyzing  the
high-resolution, high S/N spectra of four red clump stars, finding
an average [Fe/H] = $+0.47\pm 0.04\pm 0.08$, where the two error
values represent the random term and the systematic one, respectively.

This result has been obtained using spectrum synthesis of Fe {\sc i} and  Fe
{\sc ii} lines, and adopting atmospheric parameters from the photometry. The
soundness of our analysis has also been checked relatively to a well studied,
metal-rich field giant star, $\mu$ Leo.

We have measured O and C abundances using synthesis of the [O {\sc i}]
line at 6300.31 \AA \ and of the C$_2$ molecular feature at 5086 \AA,
finding [O/Fe] = $-0.32$ and [C/Fe] = $-0.23$ dex, on average, for \amm.

\amm \ represents a challenge to any model for the formation of the Galactic
disk, a possible missing link between globular and open clusters, a
genuine puzzle that modern abundance analysis based on high resolution
spectroscopy will help to unveil.

\acknowledgements

We thank the TNG staff who helped collecting the spectra. We gratefully
acknowledge the use of the BDA, created and updated by J.-C. Mermilliod for
many years.  This project has been partially supported by the Italian MIUR,
under PRIN 2003029437.

\clearpage

\begin{deluxetable}{ccccccccc}
\tablewidth{0pt}
\tablecaption{Information on the four observed stars.}
\tablehead{
\colhead{ID}         & \colhead{ID}      &
\colhead{RA}         & \colhead{Dec}  &
\colhead{V}          & \colhead{$B-V$}	&
\colhead{$K$}          & \colhead{RV} &
\colhead{S/N}\\
\colhead{(FLJ)}      & \colhead{(BDA)} &
\colhead{(2000)}     & \colhead{(2000)}&
\colhead{}           &\colhead{}      &
\colhead{2MASS}      &\colhead{km s$^{-1}$} &}
\startdata
    2014 & 2562  &19 21 01.1 &37 46 39.6   &14.563 &1.403 &11.468 & $-48.60$ 
& 60  \\
    3009 & 3363  &19 20 56.5 &37 44 33.7   &14.638 &1.387 &11.556 & $-48.26$ 
& 50  \\
    3019 & 3328  &19 20 56.2 &37 43 07.7   &14.675 &1.334 &11.586 & $-46.10$ 
& 85  \\
    SE49 & 3926  &19 20 53.1 &37 45 33.4   &14.540 &1.346 &11.513 & $-45.63$ 
& 40  \\
\enddata
\tablerefs{
FLJ: \cite{friel89}; BDA: \cite{merm95}; $B, ~B-V$: 
\cite{montgomery94}; K: \cite{cutri03}; 
RV: this paper
}
\label{tab1}
\end{deluxetable}

\clearpage
\begin{deluxetable}{lcccccccc}
\tabletypesize{\small}\tablewidth{0pt}
\tablecaption{Abundances from individual Fe lines}
\tablehead{
\colhead{ Wavel.}         & \colhead{EP}      &
\colhead{$\log gf$}         & \colhead{$\mu$~Leo}  &
\colhead{2014}          & \colhead{3009}	&
\colhead{3019}          & \colhead{SE49} &
\colhead{Mean}\\
\colhead{(\AA)}      & \colhead{(eV)} &
\colhead{}           & \colhead{}     &
\colhead{}           &\colhead{}      &
\colhead{}           &\colhead{ }}
\startdata
\multicolumn{9}{c}{Fe I}\\
 5560.22 & 4.43 & $-$1.10 & 7.71 &      & 8.00 & 8.06 & 7.91 & $7.99\pm
0.04$ \\
 5577.03 & 5.03 & $-$1.49 & 7.76 & 7.91 & 8.24 & 7.93 & 7.86 & $7.98\pm
0.08$ \\
 5618.64 & 4.21 & $-$1.34 & 7.67 & 7.74 & 8.01 & 8.02 & 8.07 & $7.96\pm
0.08$ \\
 5619.61 & 4.39 & $-$1.49 & 7.70 & 8.09 & 8.19 & 7.87 & 8.04 & $8.05\pm
0.06$ \\
 5650.00 & 5.10 & $-$0.80 & 8.04 & 8.12 & 8.17 & 8.27 & 7.81 & $8.09\pm
0.10$ \\
 5651.48 & 4.47 & $-$1.79 & 7.98 & 8.03 & 8.37 & 7.88 & 8.01 & $8.07\pm
0.10$ \\
 5661.35 & 4.28 & $-$1.83 & 8.04 & 8.03 & 7.83 & 8.13 & 8.03 & $8.01\pm
0.06$ \\
 5784.67 & 3.40 & $-$2.53 & 7.84 & 7.95 & 7.91 & 7.88 & 7.87 & $7.90\pm
0.02$ \\
 6078.50 & 4.79 & $-$0.38 & 8.02 & 7.74 & 8.21 & 7.85 & 8.17 & $7.99\pm
0.12$ \\
 6079.02 & 4.65 & $-$0.97 & 7.86 & 7.99 & 8.19 & 8.07 & 8.15 & $8.10\pm
0.04$ \\
 6082.72 & 2.22 & $-$3.53 & 7.89 & 7.91 & 7.84 & 7.87 & 7.97 & $7.90\pm
0.03$ \\
 6089.57 & 5.02 & $-$0.87 & 7.64 & 7.93 & 7.97 & 7.87 & 7.62 & $7.85\pm
0.08$ \\
 6093.65 & 4.61 & $-$1.32 & 7.74 &      & 8.08 & 7.97 & 7.83 & $7.96\pm
0.06$ \\
 6096.67 & 3.98 & $-$1.76 & 7.86 & 8.04 & 8.00 & 7.97 & 7.81 & $7.96\pm
0.05$ \\
 6098.25 & 4.56 & $-$1.81 & 7.82 & 8.12 & 8.16 & 8.08 & 7.80 & $8.04\pm
0.08$ \\
 6137.00 & 2.20 & $-$2.91 & 7.92 & 7.96 & 8.25 & 8.04 & 7.95 & $8.05\pm
0.07$ \\
 6270.23 & 2.86 & $-$2.55 & 8.09 & 8.01 & 8.04 & 8.16 & 8.07 & $8.07\pm
0.04$ \\
 6380.75 & 4.19 & $-$1.34 & 7.92 &      & 8.16 & 8.06 & 7.87 & $8.03\pm
0.08$ \\
 6481.88 & 2.28 & $-$2.94 & 8.11 & 7.86 & 8.25 & 8.00 & 7.94 & $8.01\pm
0.08$ \\
 6498.94 & 0.96 & $-$4.66 & 7.91 & 7.87 & 8.07 & 7.86 & 7.91 & $7.93\pm
0.05$ \\
 6518.37 & 2.83 & $-$2.56 & 8.24 & 8.16 & 8.14 & 8.06 & 8.47 & $8.21\pm
0.09$ \\
 6533.94 & 4.56 & $-$1.28 & 8.10 & 7.83 & 8.39 &      & 8.24 & $8.15\pm
0.14$ \\
 6574.25 & 0.99 & $-$4.96 & 8.10 & 7.98 & 8.11 & 7.98 & 7.94 & $8.00\pm
0.04$ \\
 6581.22 & 1.48 & $-$4.68 & 7.98 & 7.77 & 8.05 & 7.89 & 7.97 & $7.92\pm
0.06$ \\
 6609.12 & 2.56 & $-$2.65 & 8.11 & 7.96 & 7.95 & 8.05 & 8.24 & $8.05\pm
0.06$ \\
 6633.76 & 4.56 & $-$0.81 & 8.07 & 7.94 & 8.00 & 7.86 & 8.27 & $8.02\pm
0.09$ \\
 6713.75 & 4.79 & $-$1.41 & 7.93 & 7.77 & 8.01 & 7.98 & 8.20 & $7.99\pm
0.08$ \\
 6725.36 & 4.10 & $-$2.21 & 7.83 & 7.73 & 8.18 & 7.98 & 8.22 & $8.03\pm
0.11$ \\
 6726.67 & 4.61 & $-$1.05 & 7.90 & 7.94 & 8.19 & 8.02 & 8.05 & $8.05\pm
0.06$ \\
\multicolumn{9}{c}{Fe II}\\
 6247.56 & 3.87 & $-$2.32 & 7.74 & 8.01 & 7.55 &      & 8.50 & $8.02\pm
0.24$ \\
 6369.46 & 3.89 & $-$4.21 & 7.80 & 8.02 & 7.75 & 7.93 & 7.69 & $7.85\pm
0.12$ \\
 6383.72 & 5.55 & $-$2.09 & 7.94 & 7.88 & 8.16 & 7.85 & 8.24 & $8.03\pm
0.10$ \\
 6416.93 & 3.89 & $-$2.70 & 7.67 & 7.94 & 8.02 & 8.05 & 7.70 & $7.93\pm
0.08$ \\
 6432.68 & 2.89 & $-$3.58 & 7.84 & 7.73 & 7.67 & 7.62 & 8.32 & $7.84\pm
0.16$ \\
 6456.39 & 3.90 & $-$2.10 & 7.75 & 7.81 & 8.07 & 7.55 & 8.12 & $7.89\pm
0.13$ \\
\enddata
\label{t:abundances}
\end{deluxetable}

\clearpage

\begin{deluxetable}{lccccccccccc}
\tablewidth{0pt}
\tabletypesize{\small}
\tablecaption{Summary of atmospheric parameters and abundance derivations; 
$v_t$
is 1.05 km~s$^{-1}$ for all stars. Last line gives the cluster
mean iron abundance.}
\tablehead{
\colhead{Star}            & \colhead{$T_{\rm eff}$}           & 
\colhead{$\log g $}       & \colhead{[A/H]} & 
\multicolumn{3}{c}{$\log n$(Fe {\sc i})}        	      &
\multicolumn{3}{c}{$\log n$(Fe {\sc ii})}       	      & 
\colhead{[Fe/H]{\sc i}}   & \colhead{[Fe/H]{\sc ii}}\\
\colhead{}                & \colhead{(K)}       	      &  
\colhead{}                &   \colhead{}        	      &
\colhead{n}     & \colhead{mean}  &
\colhead{r.m.s.}   	  & \colhead{n}     & \colhead{mean}  &
\colhead{r.m.s.}   	  &  \colhead{}     & \colhead{}  }
\startdata
$\mu$~ Leo & 4490 & 2.30 & 0.47 & 29 & 7.92 & 0.15 & 6 & 7.79 &
0.10 & +0.38 & +0.30 \\
\\
2014 & 4463 & 2.30 & 0.50 &  26 & 7.94 & 0.12 & 6 & 7.90 & 0.11 &
+0.40 & +0.41 \\
3009 & 4473 & 2.33 & 0.50 &  29 & 8.10 & 0.14 & 6 & 7.87 & 0.24 &
+0.56 & +0.38 \\
3019 & 4468 & 2.35 & 0.50 &  28 & 7.99 & 0.11 & 5 & 7.80 & 0.21 &
+0.45 & +0.36 \\
SE49 & 4512 & 2.32 & 0.50 &  29 & 8.01 & 0.18 & 6 & 8.09 & 0.33 &
+0.47 & +0.60 \\
\\
$\langle$cluster$\rangle$ &       &      &      &    & 8.01 & 0.07 &   
& 7.92 & 0.08 & +0.47 & +0.43 \\
\enddata
\label{t:summary}
\end{deluxetable}

\clearpage

\begin{deluxetable}{llcclcc}
\tablewidth{0pt}
\tablecaption{Sensitivity and errors due to uncertainties in the atmospheric
parameters}
\tablehead{
\colhead{Parameter}             & \colhead{Change}                 & 
\colhead{$\Delta$[Fe/H{\sc i}]} & \colhead{$\Delta$[Fe/H]{\sc ii}} & 
\colhead{Error}                 & \colhead{[Fe/H]{\sc i}}          &
\colhead{[Fe/H]{\sc ii}} }
\startdata
$T_{\rm eff}$  & +100 K     	  & +0.016   & $-$0.158  & 100~K   & +0.016    &
$-$0.158 \\
$\log{g}$      & +0.3 dex   	  & +0.036   & +0.184	 & 0.2~dex   & +0.024    &
+0.123 \\
${\rm [A/H]}$  & +0.1 dex   	  & +0.024   & +0.044	 & 0.06~dex  & +0.014    &
+0.026 \\
$v_t$          & +0.2 km s$^{-1}$ & $-$0.124 & $-$0.047  & 0.13 km s$^{-1}$ & $-$0.075  &
$-$0.028 \\
Fitting error  &            	  &	     &  	 & 0.16~dex  & +0.026    & 
+0.099 \\
\\
total          &            	  &	     &  	 &	 &  0.086    & 
0.226 \\
\enddata
\tablecomments{Fitting error is the line-to-line uncertainty related to the procedure of
attributing an abundance to each individual line. It includes random errors due
to the photometric uncertainties on the line profile and to the estimate of the
local continuum level.}
\label{t:errors}
\end{deluxetable}

\clearpage

\clearpage

\clearpage

\clearpage

\end{document}